\documentclass [12pt]{article}
\usepackage[centertags]{amsmath}
\usepackage{amsfonts} \usepackage{amssymb}\usepackage{eufrak}
\usepackage{graphicx}
\usepackage{lscape}
\newcommand{\COMMENTO}[1]{}
\newcommand{\uno}{\mathbb{I}}
\newcommand{\sqa}{\sqrt{\alpha'}}
\newcommand{\cE}{{\cal E}}
\newcommand{\cG}{{\cal G}}
\newcommand{\Lrpu}{{N_1}}

\begin{document}

\begin{titlepage}
\rightline{DFTT-8/2008} \vskip 3.0cm
\centerline{\LARGE \bf  A comment on discrete Kalb-Ramond field  }
\vskip .5cm
\centerline{\LARGE \bf on orientifold and rank reduction.}
\vskip .5cm
%\vskip .5cm\centerline{\LARGE \bf in String Theory }
\vskip 1.0cm \centerline{\bf I. Pesando}
\vskip .6cm 
\centerline{ \sl  Dipartimento di Fisica Teorica,
  Universit\`a di Torino and INFN, Sezione di Torino,} \centerline{\sl
 via P. Giuria 1,  I-10125, Torino, Italy}
 \vskip 1cm

\begin{abstract}
We show that the rank reduction of the gauge group 
on orientifolds in presence of non
vanishing discrete Kalb-Ramond field can be explained by the presence
of an induced field strength in a non trivial bundle on the branes.
This field strength is also necessary for the tadpole cancellation and
the number of branes is left unchanged by the presence of the discrete
Kalb-Ramond background.
\end{abstract}
\COMMENTO{
\vfill  {\small{ Work partially supported by the European
Community's Human Potential Programme under contract
MRTN-CT-2004-005104 ``Constituents, Fundamental Forces and Symmetries
of the Universe'' and by the Italian M.I.U.R. under contract
PRIN-2005023102 ``Strings, $D$-branes and Gauge Theories''. }}
}
\end{titlepage}

\newpage

\section{Introduction.}
%\section{A first quick look to the problem.}
Since its discovery in (\cite{Bianchi:1991eu}) the phenomenon of rank
reduction of gauge group in presence of a non vanishing $B$ in an
orientifold has been revisited a certain number of times ( see for example
(\cite{Bianchi:1997rf}, \cite{Witten:1997bs},
\cite{Angelantonj:1999jh}),
\cite{Angelantonj:1999xf}).
We will show that there are still some points which are worth
understanding;
in particular we will show that the rank reduction 
can be understood as an effect of
an induced non trivial gauge bundle on the compact part of branes.
 
Even in presence of $B_{i j}$ 
the usual way used to describe the condition of no momentum flow
through the ends of the open string is $p_{L i}=-p_{R i}$
(\cite{Angelantonj:2002ct},\cite{Angelantonj:1999jh} and
\cite{Bianchi:1997rf}),  this can be better rewritten as
\begin{equation}
\left(p_{L i}+p_{R i}\right) | B \rangle=
\left( n_i - B_{i j} m^j\right) | B \rangle
=0
\label{naive-refl}
\end{equation} 
with $|B\rangle$ the boundary state.
On the other side the reflection condition for a generic constant
background on a torus can
be written as (see for example (\cite{DiVecchia:2007dh}, \cite{Di
  Vecchia:2006gg}, \cite{Pesando:2005df})
\begin{equation}
 \left( n_i - \hat F_{i j} m^j \right) | B(E,F) \rangle
=0
\label{proper-refl}
\end{equation} 
where $\hat F_{i j}= 2\pi \alpha' q F_{i j}$ is the adimensional field strength.
Comparing these two equations (\ref{naive-refl}) and
(\ref{proper-refl})  we find
\begin{equation}
\hat F_{i j} = B_{i j} 
\label{naive-equality}
\end{equation}
This equality is partially naive because the interesting values of $B$
are fractional and therefore the field strength cannot be the field
strength of a trivial $U(1)$ bundle on a torus.
The previous equation (\ref{naive-equality}) is naive because does not
seem to consider the effect of the worldsheet parity $\Omega$ on
the stacks of branes where we need to define the bundle.
Anyhow it is a clear hint that something is missing.
 
To make this equality  more precise we exam once again the different
amplitudes involved in the orientifold projection 
paying particular attention to the dependence on the metric, either
open or closed. 
This will essentially confirm eq. (\ref{naive-equality}), more
precisely we find a field strength 
\begin{equation}
2\pi \alpha' q F= \frac{1}{2} B_{i j} d x^i\, d x^j\, \uno_{2^{D/2}}
\end{equation}
of a $SO(2^{D/2})$ bundle along with the proper transition functions
(as in eq. (\ref{trans-func-B12}) when we write $ \uno_{2^{D/2}}=
\uno_{2^{D/2-1}}\oplus \uno_{2^{D/2-1}}$).  
This turns out not to be the only new ingredient: in order to be able
to cancel the tadpole we discover that we need some extra signs due to the
(momentum dependent) Chan-Paton matrices which naturally emerge in the
description of the string on a non trivial bundle (\cite{DiVecchia:2007dh}).
Moreover in order to get a perfect match of the open and closed string
computations we find that a phase quadratic in winding is necessary in
the definition of the crosscap state.
The non trivial bundle then leads naturally to rank reduction because
of a mechanism \`a la  Scherk-Schwarz.

The paper is organized as follows.
In section \ref{review-sect} we fix our conventions and 
review the description of an open string in a non trivial bundle on a
torus and we discuss how the effect of a non trivial bundlecan be
interpreted to give a reduction  \`a la Scherk-Schwarz. 
Then in section \ref{Omega-sect} we discuss the action of the
worldsheet parity $\Omega$ on the states in a non trivial bundle.
In section \ref{Klein-sect} we perform the usual Klein bottle computation
paying attention to the dependence on the metric of the result and we
state the final form of the crosscap, momentum dependent signs included.
In sections \ref{ann-sect} and \ref{moebius-sect} we perform the open
string computation and we derive the results (almost) correctly
guessed in the previous literature. 
Finally in section 7 we draw our conclusions.

\section{A short review of open string on non trivial bundles.}
\label{review-sect}
In the following we use the notations used in
(\cite{DiVecchia:2007dh}, \cite{Di Vecchia:2006gg}).
In particular the closed string Hamiltonian in a metric background
$E_{i j}=G_{i j}+B_{i j}$ on a generic flat space $R^{D-d}\otimes T^d$ 
 ($D=26$) can be written as follows:
\begin{eqnarray}
\frac{H_c-4}{2} 
&=&
 L_0 + \tilde L_0=
 \frac{\alpha'}{4 \pi} \int_{0}^{\pi} d \sigma \left[P_{L}^{2}
  + P_{R}^{2} \right],
\nonumber\\
&=& 
N + {\tilde{N}} 
+ \frac{1}{2}
\left[G_{i j} {\hat{m}}^i {\hat{m}}^j
  + ( {\hat{n}}_i - B_{i k} {\hat{m}}^{k} ) G^{i j}
( {\hat{n}}_j - B_{j h} {\hat{m}}^{h} )  \right]
+ \frac{\alpha'}{2} G^{\mu\nu} k_\mu k_\nu
\nonumber\\
\label{hamilto89}
\end{eqnarray}
where 
$i,j,\dots=1,2,\dots d~~;~~ \mu,\nu=0,d+1,\dots D$
and 
the explicit expressions of $L_0$ and
${\tilde{L}}_0$ are given by
\begin{eqnarray}
&&L_0 = 
\frac{\alpha'}{4} G^{\mu\nu} k_\mu k_\nu
+\frac{\alpha' }{4} G_{i j} p_{R}^i  p_{R}^j+N
~~;~~ 
N=\sum_{n=1}^{\infty} 
G_{\mu\nu} \alpha_{-n}^{\mu} \alpha_{n}^{\nu}
+G_{i j} \alpha_{-n}^{i} \alpha_{n}^{j}
\nonumber\\
&&{\tilde{L}}_0 = 
\frac{\alpha'}{4} G^{\mu\nu} k_\mu k_\nu
+\frac{\alpha' }{4} G_{i j} p_{L}^i  p_{L}^j +
\tilde N
~~;~~ 
\tilde N=
\sum_{n=1}^{\infty} 
G_{\mu\nu} {\tilde{\alpha}}_{-n}^{\mu} {\tilde{\alpha}}_{n}^{\nu}
+G_{i j} {\tilde{\alpha}}_{-n}^{i } {\tilde{\alpha}}_{n}^{j }
,\label{L078}
\end{eqnarray}
and the spectrum of the compact momenta  by
\begin{eqnarray}
(k_{i})_{\left(\begin{array}{c} L \\
                 R \end{array} \right)} =
G_{i j}(p^{j})_{\left(\begin{array}{c} L \\
                 R \end{array} \right)} =
\frac{1}{\sqrt{\alpha'}} \left[ (n_i - B_{i j}m^j )\pm G_{i j} m^j \right] ,
\label{pLpR89}
\end{eqnarray}
The non vanishing commutators are
\begin{equation}
[x_L^i, p_L^j]=i G^{i,j}
~~,~~
 [ \alpha_{L n}^{i} , \alpha_{L m}^{j}] = n \delta_{n+m,0} G^{i,j}
\end{equation}
and similarly for the right movers and for the non compact directions.
The normalization of the zero modes is
\begin{equation}
\langle k_\mu, n_i,m^i | k_\mu', n_i',m^{i'} \rangle= 
(2\pi)^{D-d} \delta^{D-d}(k_\mu-k_\mu')~
(2\pi\sqa)^d \delta_{n,n'} \delta_{m,m'}
~.
\end{equation}

Let us now consider the open string
in a metric background given by $E_{i j}%=G_{i j}+B_{i j}
$
and in presence of a constant background field $F_{i j}$.
We assume that this background field is  
\begin{eqnarray}
\hat F_{2 a-1, 2 a}= 2 \pi \alpha' q F _{2 a-1, 2 a}=
\frac{f_a}{L_{(a)}} \uno_{\prod_{b=1}^r L_{(b)} \Lrpu}
~~,~~
a=1,\dots r
\label{gen-F}
\end{eqnarray}
with all the other components vanishing, i.e. the rank of the field
strength is $r$.
This background field is obtained from the gauge field (up to gauge choices)
\begin{equation}
A_i = \frac{1}{2}F_{j i} x^j \uno_{\prod_{b=1}^r L_{(b)} \Lrpu }
\label{gen-A}
\end{equation}
along with the transition functions which we take to be
\begin{eqnarray}
\Omega_1= e^{i 2\pi \theta_1} 
e^{-i \frac{f_{1}}{L_{(1)}}  \frac{x^2}{2\sqa}}
Q_{L_1}\otimes \uno_{L_2} \dots \uno_{\Lrpu} 
&~~,~~&
\Omega_2= e^{i 2\pi \theta_2} 
e^{i \frac{f_{1}}{L_{(1)}}  \frac{x^1}{2\sqa}}
P_{L_1}^{-f_1}\otimes \uno_{L_2} \dots \uno_{\Lrpu} 
\nonumber\\
\Omega_3= e^{i 2\pi \theta_3} \uno_{L_1} \otimes 
e^{-i \frac{f_{2}}{L_{(2)}}  \frac{x^4}{2\sqa}}
Q_{L_2} \dots \uno_{\Lrpu} 
&~~,~~&
\Omega_4= e^{i 2\pi \theta_4} \uno_{L_1} \otimes
e^{i \frac{f_{2}}{L_{(2)}}  \frac{x^3}{2\sqa}}P_{L_2}^{-f_2}  \dots \uno_{\Lrpu} 
\nonumber\\
&\vdots&
\nonumber\\
\Omega_{2r+1}= e^{i 2\pi \theta_{2r+1}} \uno_{L_1} \otimes \uno_{L_2} \dots \uno_{\Lrpu} 
&~~,\dots~~&
\Omega_d= e^{i 2\pi \theta_d} \uno_{L_1} \otimes
\uno_{L_2}  \dots \uno_{\Lrpu} 
\label{gen-Omega}
\end{eqnarray}
where $e^{i 2\pi \theta_i}$ are the abelian Wilson lines.
The block diagonal field strength (\ref{gen-F}), the fact we are working on a 
torus $T^d$ %$\prod_{a=1}^r T^2_{(a)} \otimes T^{d-2r}$
and the cocycle conditions on the transition functions, i.e
$\Omega_j(x^k+2\pi \sqa \delta^k_i)\Omega_i(x^k)=
\Omega_i(x^k+2\pi \sqa \delta^k_j)\Omega_j(x^k)
$, oblige to
consider the gauge group to be factorized as
$\otimes_{a=1}^{r} U(L_{(a)})\otimes U(\Lrpu)$: this does {\sl not } happen on a
non compact surface and it is responsible for the rank reduction as we
can see from the physical states in eq. (\ref{op-str-phy-st}) and we
discuss after eq. (\ref{exam-comp-norm}). 
\COMMENTO{
\footnote{
We consider tensor product of bundles and not direct sums because
a field strength proportional to the unity can always be rotated
in a block diagonal form and hence we can consider factorized torii.
Because of this the transition matrices in two directions on two
different torii must commute and this implies that the bundle is
a tensor product.\newcommand{\cE}{{\cal E}}

} on a factorized torus $\prod_{a=1}^r T^2_{(a)} \otimes T^{d-2r}$.
}

On this background the dipole string Hamiltonian  is then given by
\begin{eqnarray}
H_o-1=L_0 
&=& 
\alpha' G^{\mu\nu} k_\mu k_\nu
+\alpha'  p^i {\cal{G}}_{i j} p^j 
+  \sum_{n=1}^{\infty} n 
{{G}}_{\mu \nu} \alpha^{\dagger \mu}_{n} \alpha^{\nu}_{n} 
+{\cal{G}}_{i j} a^{\dagger i}_{n} a^{j}_{n} 
\nonumber\\
&=&
\alpha' G^{\mu\nu} k_\mu k_\nu
+ 
{\cal{G}}^{i j}  
\frac{n_i}{L_i} \frac{n_j}{L_j}
%n_i n_j 
+ \sum_{n=1}^{\infty}  
{{G}}_{\mu \nu} \alpha^{\dagger \mu}_{n} \alpha^{\nu}_{n} 
+{\cal{G}}_{i j} a^{\dagger i}_{n} a^{j}_{n} 
\label{L061}
\end{eqnarray}
with the open string metric given by
\begin{eqnarray}
{\cal{G}}_{i j} =  G_{i j} - {\cal B}_{i k} G^{k h} {\cal B}_{h j} =
{\cal{E}}^{T}_{i k} G^{k h} {\cal{E}}_{h j}
\label{openme2}
\end{eqnarray}
In the previous expression we have defined the following quantities
\begin{eqnarray}
{\cal{B}}_{i j} 
&=& 
B_{i j} - \hat F_{i j}
\label{calB}
\\
{\cal{E}}_{i j} 
&=& 
G_{i j} -{\cal{B}}_{i j} 
= G_{i j}
- B_{i j} + \hat F_{i j}
\label{calE}
\end{eqnarray}
The compact momenta  have spectrum
\begin{eqnarray}
%k_i = \frac{n_i}{\sqrt{\alpha'}W^i} \Longrightarrow 
%p^i = {\cal{G}}^{i j} \frac{n_j}{\sqrt{\alpha'} } 
p^i = {\cal{G}}^{i j} \frac{1}{\sqa} \frac{n_i}{L_i}
\label{impu43}
\end{eqnarray}
where we have defined $L_{ 2 a}= L_{2 a -1}=L_{(a)}$ for $1\le a \le r$, $L_i=1$ for $2 r
< i \le d$.
The non-vanishing commutation relations in compact directions are:
\begin{eqnarray}
[ x^i , p^j ] = i {\cal{G}}^{i j}
~~;~~
[ x^i , x^j ] = i ~2\pi \alpha' \Theta^{i j}
~~;~~ 
[ \alpha_{n}^{i} , \alpha_{m}^{j}] = n \delta_{n+m,0}
{\cal{G}}^{i j}
\label{comure}
\end{eqnarray}
where $\cE^{-1}= \cG^{-1} - \Theta$.
On this background the normalized  string states are given by
\begin{eqnarray}
|\chi; k_\mu, n_i; u\rangle
&=&
\frac{1}{(2\pi\sqa)^{d/2}}
|\chi \rangle \otimes
|k_\mu\rangle 
\otimes 
\Lambda_{L_{(1)}; I_1 J_1}(n_1,n_2) 
~|\frac{n_1}{\sqa L_{(1)}},\frac{n_2}{\sqa L_{(1)}} \rangle_p
~|J_1 I_1\rangle
\nonumber\\
&&
\otimes 
\Lambda_{L_{(2)}; I_2 J_2}(n_3,n_4)
~|\frac{n_3}{\sqa L_{(2)}},\frac{n_4}{\sqa L_{(2)}} \rangle_p
~ |J_2 I_2\rangle
\dots
\nonumber\\
&&
\otimes
~T_{u~\Lrpu; I_{r+1} J_{r+1}}
~| \frac{ n_{2 r+1}}{\sqa},\dots \frac{ n_{d}}{\sqa}\rangle 
~|J_{r+1} I_{r+1}\rangle
\nonumber\\
\label{op-str-phy-st}
\end{eqnarray} 
where $|\chi \rangle$ is the collective name for the quantum numbers
associated with the non zero modes,
$|\frac{n_1}{\sqa L_{(1)}},\frac{n_2}{\sqa L_{(1)}} \rangle_p$ is a momentum
eigenvector,
$|J_1 I_1\rangle$ is an element of basis for the color indexes (see
(\cite{DiVecchia:2005vm}) for more details). 
The meaning of writing $\Lambda_{L_{(1)}; I_1 J_1}(n_1,n_2) ~|J_1 I_1\rangle$
is that for a given momentum 
$\left(\frac{n_1}{\sqa L_{(1)}},\frac{n_2}{\sqa L_{(1)}}\right)$
not all the possible $L_{(1)}^2$  $~|J_1 I_1\rangle$ color index
combinations are possible, as it is usual with the trivial bundle, but only one.

In the eq. (\ref{op-str-phy-st}) 
$T_u$ are the usual $\Lrpu^2$ hermitian $u(\Lrpu)$ generators 
and can be traded for the $\Lrpu^2$ color states $|J_{r+1} I_{r+1}\rangle$.

The $\Lambda$ are
the hermitian momentum dependent Chan-Paton matrices given by
\begin{eqnarray}
~\Lambda_{L; I J}(n_1,n_2)
&=&
\frac{1}{\sqrt{L}} e^{-i \frac{\pi}{L} \hat h n_1 n_2}
\left( Q_L^{\hat h n_2} P_L^{-n_1}\right)_{I J}
~~,~~
0\le I,J < L
\end{eqnarray}
with $\hat h f \equiv -1 ~~ mod~L$ which enjoy the hermitian
conjugation property
\begin{equation}
\Lambda_L^\dagger(n_1,n_2)=\Lambda_L(-n_1,-n_2)
\end{equation}
and are normalized as
\begin{equation}
tr\left(\Lambda_L^\dagger(n_1,n_2) \Lambda_L(m_1,m_2) \right)
= \delta_{n,m}
\end{equation}
In particular for the $L=2$ case, which is of our interest, the explicit
form of the $\Lambda_2$ matrices is
\begin{equation}
~\Lambda_{2; I J}(n_1,n_2)
=
\frac{1}{\sqrt{2}} e^{-i \frac{\pi}{2}  n_1 n_2}
\left( \sigma_3^{ n_2} \sigma_1^{-n_1}\right)_{I J}
~~,~~
0\le I,J < 2
\end{equation}
To show that the states in eq.  (\ref{op-str-phy-st}) are normalized
we perform a computation like 
\begin{eqnarray}
&&
\langle M K |
~{}_p\langle\frac{n_1}{\sqa L},\frac{n_2}{\sqa L} |
~\left(\Lambda^\dagger_{L}\right)_{ K M}(n_1,n_2) 
~\Lambda_{L; I J}(m_1,m_2) 
~|\frac{m_1}{\sqa L},\frac{m_2}{\sqa L} \rangle_p
~|J I\rangle
\nonumber\\
&&
=
(2\pi\sqa)^2 \delta_{n,m}
~ \delta_{K,I}\delta_{M,J}
~\left(\Lambda^\dagger_{L}\right)_{ K M}(n_1,n_2) 
~\Lambda_{L; I J}(m_1,m_2) 
\nonumber\\
&&
=
(2\pi\sqa)^2 \delta_{n,m}
~tr\left(
\Lambda_{L}^\dagger(n_1,n_2) 
~\Lambda_{L}(m_1,m_2)
\right)
=
(2\pi\sqa)^2 \delta_{n,m}
\label{exam-comp-norm}
\end{eqnarray}

Finally we notice that even if we start with $\prod_{b=1}^r L_{(b)} \Lrpu$
branes the number of massless states $k_\mu^2=0$ is only $\Lrpu^2$. 
This does not mean that we have not $(\prod_{b=1}^r L_{(b)} \Lrpu)^2$
states as we naively would expect but that some of them become
massive, with a mass of order $\frac{1}{L}$:
it is essentially a Scherk-Schwarz reduction mechanism
and is the key idea of the explanation of the rank reduction.

For later convenience and use in the computation of the annulus and the
Moebius amplitude we write the spectral decomposition of the unity
as%
\begin{eqnarray}
\uno 
&=&
\int \frac{d^{D-d}k_\mu}{ (2\pi)^{D-d}}
\sum_{\chi,n_i,u}
|\chi; k_\mu, n_i; u\rangle
%\frac{1}{(2\pi\sqa)^d}
~\langle \chi; k_\mu, n_i; u|
\nonumber\\
&=&
\int \frac{d^{D-d}k_\mu}{ (2\pi)^{D-d}}
\sum_{\chi,n_i,u}
|\chi \rangle \otimes
|k_\mu\rangle 
\otimes 
\Lambda_{L_{(1)}; I_1 J_1}(n_1,n_2) 
~|\frac{n_1}{\sqa L_{(1)}},\frac{n_2}{\sqa L_{(1)}} \rangle_p
~|J_1 I_1\rangle
\nonumber\\
&&
\hspace{6em}
\frac{1}{(2\pi\sqa)^d}
\dots
\langle M_1 K_1|
~{}_p\langle\frac{n_1}{\sqa L_{(1)}},\frac{n_2}{\sqa L_{(1)}} |
~\Lambda^\dagger_{L_{(1)}; K_1 M_1}(n_1,n_2) 
\otimes
\langle k_\mu|
\otimes 
\langle\chi|
\nonumber\\
\end{eqnarray}
which can be used to define the trace as
\begin{eqnarray}
Tr(O) 
&=&
\int \frac{d^{D-d}k_\mu}{ (2\pi)^{D-d}}
\sum_{\chi,n_i,u}
%\frac{1}{(2\pi\sqa)^d}
~\langle \chi; k_\mu, n_i; u|
~O
|\chi; k_\mu, n_i; u\rangle
\end{eqnarray}

\section{The action of $\Omega$ on a non trivial bundle.}
\label{Omega-sect}
Before computing the open amplitudes we must discuss the action of
the $\Omega$ operator.
Our picture is to start with a stack of $\prod_{b=1}^r L_{(b)} \Lrpu$
branes  which gets mapped to a second image stack of $\prod_{b=1}^r L_{(b)}
\Lrpu$ branes by the $\Omega$ action therefore the generic element for
the color basis will be
\begin{equation}
|C D \rangle
=\left( \begin{array}{cc}
| c d \rangle & | c d' \rangle \\
| c' d \rangle & | c' d' \rangle\\
\end{array}
\right)
\label{2x2-col-ind}
\end{equation}
where we have used the usual convention of using the prime to denote
the color indexes of the mirror stack and the color index 
$c$ is the collective name of the indexes $(I_1 ~I_2\dots I_{r+1})$.

On the fluctuations around the trivial background we use the usual
action 
\begin{equation}
\Omega |\chi, k_i=\frac{n_i + \theta_{C i}- \theta_{D i} }{\sqa} ; C D\rangle = 
(\gamma_\Omega)_{C D_1} 
| \chi^\Omega,k_i= \frac{n_i + \theta_{C_1 i}- \theta_{D_1 i} }{\sqa};
C_1 D_1\rangle 
(\gamma_\Omega^{-1})_{C_1 D}
\label{Omega-ph-st-trivial}
\end{equation}
where $C,D,\dots=1,\dots 2\prod_{b=1}^r L_{(b)} \Lrpu$,
 $\theta_{C}$ is the Wilson line on the $C$-th brane and we
assume that $\gamma_\Omega$ have the simplest off diagonal form
\begin{equation}
\gamma_\Omega
=\left( \begin{array}{cc}
& \uno \\
\uno & \\
\end{array}
\right)
\end{equation}
which is the best suited form to describe the picture where we start
with a stack of $\prod_{b=1}^r L_{(b)} \Lrpu$ branes which get mapped to a new stack.
The $\Omega$ acts on the non zero modes as 
\begin{equation}
\Omega \alpha_n \Omega = (-)^n \alpha_n
\end{equation}
Therefore the gluons which survive the projection are those 
whose Chan Paton matrices $\Lambda$ satisfy 
$\gamma_\Omega \Lambda \gamma_\Omega^{-1}=-\Lambda$ which with our
choice of $\gamma_\Omega$ reads\footnote{
To obtain the $so(2\prod_{b=1}^r L_{(b)} \Lrpu)$ matrices we have to choose 
$\gamma'_\Omega
=\left( \begin{array}{cc}
\uno & \\
& \uno  \\
\end{array}
\right) $, 
choice which is anyhow equivalent to $\gamma_\Omega$ since
$\gamma'_\Omega = -i U ^\dagger \gamma_\Omega U^*$ where
$U= \frac{1}{\sqrt{2}}
\left( \begin{array}{cc}
\uno & i \uno\\
i \uno & \uno  \\
\end{array}
\right)
$.
}
\begin{equation}
\Lambda
=\left( \begin{array}{cc}
H & A \\
A^\dagger & -H^T \\
\end{array}
\right)
~~,~~
A^T=-A,~~ H^\dagger=H
\end{equation}
This form of an element of the algebra suggests that if we start
with the transition functions $\Omega_{(1) i}$ on the original stack
the transition functions for the two stacks are given by
\begin{eqnarray}
\Omega_i
&=&
\left(\begin{array}{c c}
\Omega_{(1) i} & \\
& \Omega_{(1) i}^* 
\end{array} \right)
\label{trans-func-B0}
\end{eqnarray}
This form is consistent with the naive expectation that if the first
stack has common Wilson lines described by
\begin{equation}
\Omega_{(1) i}= e^{i 2\pi \theta_i} \uno
\label{common-wilson-1}
\end{equation}
the Wilson lines on the image stack are opposite so that the strings
which connect the two stacks have double Wilson line $\pm 2 \theta_i$
which corresponds to halving the T-dual torus.

If we now consider a non trivial bundle of the kind described in the
previous section when $B=0$ we can still take the transition functions as in
eq. (\ref{trans-func-B0}).
Since in the transition functions it is essentially encoded the
background field, see  eq. (\ref{gen-Omega}),  we deduce that 
the background field strength on the image stack  is $F^\Omega=-F$ 
because the transition functions on the image stack are
the complex conjugate of the original ones.
This is nice since $\cE^\Omega=\cE^T$ and therefore $\cG^\Omega=\cG$.

The action of $\Omega$ on the fluctuation around this non trivial
background is still in nuce given by eq. (\ref{Omega-ph-st-trivial} ) 
\begin{equation}
\Omega |\chi
%, k_i=\frac{\frac{n_i}{L_i} + \theta_{C i}- \theta_{D i}}{\sqa}
;C D\rangle = 
(\gamma_\Omega)_{C D_1} 
| \chi^\Omega
%,k_i= \frac{\frac{n_i}{L_i} + \theta_{C_1 i}- \theta_{D_1 i} }{\sqa}
; C_1 D_1\rangle 
(\gamma_\Omega^{-1})_{C_1 D}
\end{equation}
where we have not written the momenta since the strings connecting the
two stacks are dicharged ones.

If we now consider a non trivial gauge background with some half
integer $B$ turned on together with the the transition functions as in
eq. (\ref{trans-func-B0}) then we lose the nice properties
$\cE^\Omega=\cE^T$ and $\cG^\Omega=\cG$.
On the other hand, as we discuss later in section \ref{ann-sect}, 
tadpole cancellation requires that we freeze $\hat F-B=\hat
F^\Omega-B=0$, i.e. $\hat F^\Omega= - \hat F +2 B$.
This requires that we consider the transition functions
\begin{eqnarray}
\Omega_i
&=&
\left(\begin{array}{c c}
\Omega_{(1) i} & \\
& e^{-2i  B_{i j} \frac{x^j}{2 \sqa} }\Omega_{(1) i}^* 
\end{array} \right)
=
\left(\begin{array}{c c}
\Omega_{(1) i} & \\
& \Omega_{(1) i} 
\end{array} \right)
\label{trans-func-B12}
\end{eqnarray}
and this in turn implies that on both stacks we have the same field
strength so that also the strings connecting the two stacks are dipole
ones therefore the action of $\Omega$ on the fluctuation around this non trivial
background is given by 
\begin{equation}
\Omega |\chi
, k_i=\frac{\frac{n_i}{L_i} + \theta_{C i}- \theta_{D i}}{\sqa}
;C D\rangle = 
(\gamma_\Omega)_{C D_1} 
| \chi^\Omega
,k_i= \frac{\frac{n_i}{L_i} + \theta_{C_1 i}- \theta_{D_1 i} }{\sqa}
; C_1 D_1\rangle 
(\gamma_\Omega^{-1})_{C_1 D}
\end{equation}
As a consequence the spectrum is essentially 
given by four times the spectrum in eq. (\ref{op-str-phy-st});
using the the same notation as in eq. (\ref{2x2-col-ind}) and in
presence of the Wilson lines (\ref{common-wilson-1}) we have:
\begin{equation}
\left(\begin{array}{c c}
 |\chi; k_\mu, n_i; u\rangle & 
|\chi; k_\mu, n_i+ 2L_i \theta_i; u\rangle \\
|\chi; k_\mu, n_i- 2L_i \theta_i; u\rangle &
|\chi; k_\mu, n_i; u'\rangle
\end{array} \right)
\end{equation}

\section{The Klein bottle.}
\label{Klein-sect}
We start to compute the Klein bottle%
\footnote{ We use the amplitude normalizations given in the review 
(\cite{DiVecchia:2005vm}), from which 
we take also the following notations and relations:
$
f_1(q)\equiv q^{{1\over 12}} \prod_{n=1}^\infty (1 - q^{2n})
=e^{-\pi\tau/12}  \prod_{n=1}^\infty (1 - e^{-2 n \pi\tau})
$ with $q=e^{-\pi\tau}\in R$,  $f_1(e^{-\pi t})=
\frac{1}{\sqrt{t}}f_1(e^{-\frac{\pi}{t}})$ and
$f_1(i e^{-\pi t}) = \frac{1}{\sqrt{2 t}} f_1(i e^{-\frac{\pi}{4 t}})$ 
} for a space time 
$R^{D-d}\otimes T^{d}$ ($D=26$)
\begin{equation}
Z_{K}=\int \frac{d \tau}{\tau} 
~Tr_c\left( \frac{\Omega}{2} e^{-\pi \tau H_c}\right)
\end{equation}
where we have used $\tau$ as integration variable since this amplitude
must be interpreted as open channel amplitude even if derived it projecting
a closed string one with the insertion of  $ \frac{\Omega}{2}$ to implement the
orientifold (in this case the worldsheet parity only).

To perform the previous computation we need defining the action of
$\Omega$ (\cite{Bianchi:1997rf}). 

%>>>>>>>>> Now the commonlore follows<<<<<<<<<<

Since this is the worldsheet parity which exchange the left
and right sector we requite that the left momenta lattice 
$\{ \sqa ~G p_{L }(n,m)= \left( n + E^T m\right) \}$
%($ p_L = \parallel (p_L)^i  \parallel =\parallel e^{I i} (p_L)_I  \parallel$)
be equal to the right momenta lattice
$\{ \sqa ~G p_R(n,m)= \left( n - E m\right) \}$.
This means that for any $n, m$ we can find some $n', m'$ so that
$ p_L(n,m)= p_R(n',m')$, if we want this relation be valid for any
metric $G$ we find $m'=-m$ and $n'=n-2 B m$. This last relation must
be true for all $m$ and $n$ and therefore we restrict $2 B_{i j}\in Z$. 
Hence  we define the action of $\Omega$ as
\begin{eqnarray}
\Omega |n,m\rangle &=& | n -2 B m , -m\rangle
\nonumber\\
\Omega \alpha_n \Omega &=&  (-1)^n \tilde \alpha_n
\end{eqnarray}
Without loss of generality we can assume $B$ to be block diagonal in
the space indexes.

The Klein bottle amplitude is then given by
\begin{eqnarray}
Z_{K}&=&\int \frac{d \tau}{ 2 \tau} 
\int \frac{d^{D-d} k^\mu}{ (2 \pi)^{D-d}} (2\pi)^{D-d} \delta^{D-d}(k^\mu-k^\mu) 
e^{-\pi \tau  \alpha' (k^\mu)^2} 
\sum_{n\in Z^{d}} e^{ -\pi \tau n^T G^{-1} n }
\frac{ e^{4 \pi \tau} }
 { \left( \prod \left(1- e^{-4 \pi\tau n}\right)  \right)^{D-2}}
\nonumber\\
&=&
\int \frac{d \tau}{ 2 \tau}
\frac{V_{nc}}{(\alpha' \tau)^{\frac{D-d}{2}}}
\sum_{n\in Z^{d}} e^{ -\pi  \tau n^T G^{-1} n }
\frac{e^{ - \pi \tau \frac{26-D}{6} } }
{ \left( f_1(e^{-2\pi\tau})\right)^{D-2}  }
\end{eqnarray}
where the power $4$ in $e^{-4 \pi\tau n} $ is due to the left/right
identification and $V_{nc}=(2\pi)^{D-d} \delta^{D-d}(0)$. 
We can now perform the modular transformation on the $f_1$ and the
Poisson resummation to get
\begin{eqnarray}
Z_{K}&=&
\frac{V_{nc}}{ \sqrt{\alpha'}^{D-d}}
\int \frac{d \tau}{4 \tau} 
\frac{1}{\tau^{\frac{D-d}{2}}}
\left(\det \tau G^{-1} \right)^{-1/2}
\sum_{u\in Z^{d}} e^{ -\pi  \frac{1}{\tau} u^T G u }
 \frac{ (2\tau) ^{\frac{D-2}{2}} e^{ - \pi \tau \frac{26-D}{6} }
}{
\left( f_1(e^{ -\frac{\pi}{2\tau} } ) \right)^{D-2}
}
\nonumber\\
&=&
\frac{V_{nc}}{ \sqrt{\alpha'}^{D-d}}
~\sqrt{\det G}
~2^{\frac{D}{2}}
\int \frac{d \tau}{4 \tau^2 } 
\sum_{u\in Z^{d}} e^{ -\pi  \frac{1}{\tau} u^T G u }
 \frac{ e^{ - \pi \tau \frac{26-D}{6} }
}{
\left( f_1(e^{ -\frac{\pi}{2\tau} } ) \right)^{D-2}
}
\end{eqnarray}
Renaming $t=\frac{1}{2\tau}$ we can finally write in the closed channel
\begin{eqnarray}
Z_K&=&
2^{\frac{D}{2}}
~\frac{V_{nc}}{ 2 {\alpha'}^{\frac{D-d}{2}}}
~\sqrt{\det G}
\int d t 
\sum_{u\in Z^{d}} e^{ -\pi  ~2 t ~ u^T G u }
 \frac{ e^{ - \pi  \frac{1}{t} \frac{26-D}{6} }
}{
\left( f_1(e^{ -\pi t } ) \right)^{D-2}
}
\label{Klein}
\end{eqnarray}
where it is important to stress that the amplitude is proportional to
the determinant of the closed string metric $G$, fact that is
reflected in the normalization of the crosscap\footnote{
The dependence on $G$ and not on $E^T E^{-1} G$ in the non zero mode
part and the non trivial phase $e^{i 2\pi \sum_{i<j} s^i B_{i j} s^j}$
in the zero mode part can be seen from the Moebius amplitude.
}:
\begin{eqnarray}
|C(E)\rangle
&=&
\frac{T'_{25}}{2  }
2^{\frac{D}{4}} (\det { G})^{\frac{1}{4}}
|C(E)\rangle_{zm}
|C(E)\rangle_{nzm}
\nonumber\\
|C(E)\rangle_{zm}
&=&
|k_\mu=0\rangle \otimes
\sum_{s\in Z^d} e^{i \frac{\pi}{2} \sum_{i<j} m^i B_{i j} m^j}
|n_i = B_{i j} m^j, m^i=2 s^i\rangle 
\nonumber\\
|C(E)\rangle_{nzm}
&=&
e^{
- \sum_{n=1}^\infty (-)^n a^{\mu\dagger}_n G_{\mu\nu} \tilde  a^{\nu\dagger}_n
- \sum_{n=1}^\infty (-)^n a^{i\dagger}_n G_{i j} \tilde  a^{j\dagger}_n
}
|0,\tilde 0\rangle
\label{C-state}
\end{eqnarray}
which can be obtained from the closed string computation
\begin{equation}
Z_K =\langle C(E) | 
\frac{\alpha' \pi }{2} \int_0^\infty d l~ 
e^{-\pi l (L_0+\tilde L_0  -2)}
\delta_{L_0,\tilde L_0}
| C(E)\rangle
\end{equation}
up to the phase 
$e^{i \frac{\pi}{2} \sum_{i<j} m^i   B_{i j} m^j}
=
e^{i \pi \sum_{i<j} s^i   2B_{i j} s^j}
$
that can be determined from the interference term with the boundary
state, the Moebius amplitude.
We notice that this phase is half the corresponding one of the boundary state
(\ref{B-state}).

\section{The annulus amplitude.}
\label{ann-sect}
Now we want to compute annulus amplitude for open strings associated
with $2 \prod_{b=1}^r L_{(b)} \Lrpu$ branes with the gauge bundles
described in section \ref{Omega-sect}%
\footnote{
Here we are cheating a little since we know the answer.
The proper computation would be to start with $\prod_{b=1}^r L_{(b)}
\Lrpu$ branes with field strength $F$ and $\prod_{b=1}^r L_{(b)}
\Lrpu$ branes with field strength $F^\Omega$. Generically we would
have $F\ne F^\Omega$ and we would find dicharged string. In any case the
boundary state would be given by the sum of two boundary states like
(\ref{B-state}): one as in eq. (\ref{B-state}) with the substitution
$N\rightarrow \frac{N}{2}$
and one obtained from the one in eq. (\ref{B-state}) with the
substitutions
$N\rightarrow \frac{N}{2}$ and $F \rightarrow F^\Omega$.
Even in this case we would reach the same conclusion, i.e. 
$\hat F-B=\hat F^\Omega-B=0$.
}
\begin{equation}
Z_{A}=2*\int \frac{d \tau}{ 2 \tau} ~ Tr_o\left( \frac{1}{2} e^{-2 \pi \tau H_o}\right)
\label{C-step0}
\end{equation}
where the trace is taken also over the Chan Paton factors and
the factor $2$ takes into account the two possible
orientations.
Moreover the factor $\frac{1}{2}$ has been inserted into the trace as
the part of the projector $\frac{1+\Omega}{2}$ which contributes to the
annulus.

Having defined the matrix  $L= diag (L_i)$ we get therefore
\begin{eqnarray}
Z_{A}&=& 
N^2~ \int \frac{d \tau}{ 2 \tau}
\int \frac{d^{D-d} k^\mu}{ (2 \pi)^{D-d}} 
(2\pi)^{D-d} \delta^{D-d}(0) 
e^{-2\pi \tau  \alpha' (k^\mu)^2} 
\sum_{n\in Z^{d}} e^{ -2 \pi \tau n^T L^{-T} {\cal G}^{-1} L^{-1} n }
\nonumber\\
&&\times
\frac{ e^{ 2 \pi \tau} }
 { \left( \prod \left(1- e^{- 2\pi\tau n}\right)  \right)^{D-2}}
\nonumber\\
&=&
N^2
\int \frac{d \tau}{2\tau}
\frac{V_{nc}}{ (2 \alpha' \tau)^{\frac{D-d}{2}}}
\sum_{n\in Z^{d}} e^{ -2 \pi \tau n^T L^{-T}{\cal G}^{-1} L^{-1} n }
\frac{ e^{ - \pi \tau \frac{26-D}{12} }  }
{  \left( f_1(e^{-\pi\tau })\right)^{D-2} }
\label{C-step1}
\end{eqnarray}
where $N^2=(2\Lrpu)^2$ is the contribution from the Chan-Paton factors.
We can now perform the Poisson resummation and modular transformation
 and we obtain
\begin{eqnarray}
Z_{A}&=& 
N^2
\frac{V_{nc}}{ \sqrt{\alpha'}^{D-d}}
\int \frac{d \tau}{2 \tau}
\frac{1}{\tau^{D/2}}  
\left(\det 2 \tau L^{-T} {\cal G}^{-1} L^{-1}\right)^{-1/2}
\sum_{u\in Z^{d}} e^{ - \pi ~(L u)^T \frac{{\cal G}}{2 \tau} ~(L u) }
\frac{  \left(\tau \right)^{ \frac{D-2}{2} } 
        e^{ - \pi \tau \frac{26-D}{12} }  }
{  \left( f_1(e^{-\pi\frac{1}{\tau} })\right)^{D-2} }
\nonumber\\
&=&
\frac{ N^2~\det L}{ 2^{\frac{D}{2} }} 
\frac{V_{nc} ~\sqrt{ \det {\cal G} } }{2 \alpha'^{\frac{D-d}{2}}}
\int \frac{d \tau}{ \tau^2}
\sum_{u\in Z^{d}} e^{ -\pi u^T \frac{{\cal G}}{2\tau} u }
\frac{  e^{ - \pi \tau \frac{26-D}{12} }  }
{  \left( f_1(e^{-\frac{\pi}{\tau} })\right)^{D-2} }
\label{C-step2}
\end{eqnarray}
renaming $t=\frac{1}{\tau}$ we get in the closed string channel
\begin{eqnarray}
Z_A
&=&
\frac{ N^2~\det L}{ 2^{\frac{D}{2} }} 
\frac{V_{nc}  }{2 \alpha'^{\frac{D-d}{2}}}
~\sqrt{ \det {\cal G} }
\int d t
\sum_{u\in Z^{d}} e^{ -\pi \frac{t}{2} u^T {\cal G} u }
\frac{  e^{ - \pi \frac{2}{t}\frac{26-D}{12} }  }
{  \left( f_1(e^{-\pi t })\right)^{D-2} }
\label{C-closed}
\end{eqnarray}
Here the amplitude is proportional to the determinant of the open
string metric $\cal G$ which again can be seen in the normalization of
the boundary state
\begin{eqnarray}
|B(E,F)\rangle
&=&
-\frac{T'_{25}}{2}
N~2^{-\frac{D}{4}} ~\sqrt{\det L} ~(\det  {\cal G})^{\frac{1}{4}}
|B(E,F)\rangle_{zm}
|B(E,F)\rangle_{nzm}
\nonumber\\
|B(E,F)\rangle_{zm}
&=&
|k_\mu=0\rangle \otimes
\sum_s e^{ i \pi \sum_{i < j} m^i \hat F_{i j} m^j} 
|n_i = \hat F_{i j} m^j, m^i=L_i s^i\rangle 
\nonumber\\
|B(E,F)\rangle_{nzm}
&=&
e^{
- \sum_{n=1}^\infty  a^{\mu\dagger}_n (G)_{\mu\nu}\tilde a^{\nu\dagger}_n
- \sum_{n=1}^\infty  a^{i\dagger}_n (G\cE^{-1} \cE^T)_{i j}\tilde a^{j\dagger}_n
}
|0,\tilde 0\rangle
\label{B-state}
\end{eqnarray}
where we have supposed $\hat F_{i j}\propto \frac{1}{L_i}$ and the a
priori non trivial phase in the zero mode part has been derived in
(\cite{DiVecchia:2007dh},\cite{Duo:2007he}) from the path ordering and
it is necessary to allow a correct factorization of the two loop amplitude. 
Again this phase quadratic in windings cannot be seen from the closed
string computation
\begin{equation}
Z_A =\langle B(E,F) | 
\frac{\alpha' \pi }{2} \int_0^\infty d l~ 
e^{-\pi l (L_0+\tilde L_0  -2)}
\delta_{L_0,\tilde L_0}
| B(E,F)\rangle
\end{equation}
and actually in the case at hand where $L=2$ it is trivial.

Already now it is clear that if we want to cancel the
$|n=0,m=0\rangle$ state which is responsible of the tadpole
from the sum of the crosscap (\ref{C-state})
and the boundary state (\ref{B-state}) 
\begin{equation}
(|C(E)\rangle + |B(E,F)\rangle) |_{m=n=0}
=
\frac{T'_{25}}{2}
\left(
2^{\frac{D}{4}} (\det { G})^{\frac{1}{4}}
-
N~2^{-\frac{D}{4}} ~\sqrt{\det L} ~(\det  {\cal G})^{\frac{1}{4}}
\right)
|0,0\rangle
\label{tadpole}
\end{equation}
we need to have
\begin{eqnarray}
\det G &=& \det {\cal G} \Rightarrow B+\hat F=0,
\label{b+f}
\\
N ~\sqrt{\det L} &=& 2^{D/2}
\label{rk-red}
\end{eqnarray}
Now from the first equation we deduce that $L=2$ for each torus where
we have a non vanishing $B\sim \frac{1}{2}$, 
therefore we can evaluate $\det L= 2^{2
  r}$ where $r=rk(B)$ is the rank of the matrix $B$.
It then follows the usual rank reduction of the gauge group.

We can actually generalize eq. (\ref{C-step1}) to the case of the
presence of Wilson
lines. We start with a stack of $\prod_{b=1}^r L_{(b)} \Lrpu= 2^r \Lrpu$
%$N_1=\frac{N}{2}$ 
branes with field strength
$F$ and a common Wilson line $\theta_1=\theta$.
The action of $\Omega$ is to map this stack to an image stack of 
$\prod_{b=1}^r L_{(b)} N_2 %\Lrpu
=2^r N_2$
($N_2=N_1$)
branes with $F_2=F^\Omega$ and $\theta_2=-\theta$.
Therefore we have four different open string sectors which contribute to the
annulus:
\begin{eqnarray}
Z_{C}&=& 
 \int \frac{d \tau}{2 \tau}
\frac{V_{nc}}{ (2 \alpha' \tau)^{\frac{D-d}{2}}}
\Bigl[
(N_1^2+N_2^2)
\sum_{n\in Z^{d}} e^{ -2 \pi \tau n^T L^{-1} {\cal G}^{-1} L^{-1} n }
\nonumber\\
&&
+ N_1 N_2
\sum_{n\in Z^{d}} e^{ -2 \pi \tau (n+2 L \theta)^T L^{-1} {\cal G}^{-1}
  L^{-1} (n+2 L \theta) }
+ N_1 N_2
\sum_{n\in Z^{d}} e^{ -2\pi \tau (n-2 L \theta)^T L^{-1} {\cal G}^{-1}
  L^{-1} (n-2 L \theta) }
\Bigr]
\nonumber\\
&&
\hspace{6em}
\times
\frac{ e^{ - \pi \tau \frac{26-D}{12} }  }
{  \left( f_1(e^{-\pi\tau })\right)^{D-2} }
\label{C-step1-bis}
\end{eqnarray}
or in the closed string channel
\begin{eqnarray}
Z_C
&=&
\frac{\det L}{ 2^{\frac{D}{2} }} 
\frac{V_{nc}  }{2 \alpha'^{\frac{D-d}{2}}}
~\sqrt{ \det {\cal G} }
\int d t
\frac{  e^{ - \pi \frac{1}{t}\frac{26-D}{24} }  }
{  \left( f_1(e^{-\pi t })\right)^{D-2} }
\nonumber\\
&&
\sum_{u\in Z^{d}} e^{ -\pi \frac{t}{2} u^T {\cal G} u }
\left[
(N_1^2+N_2^2)
+N_1 N_2 e^{-4 i \pi u^T L \theta}
+N_1 N_2 e^{+4 i \pi u^T L \theta}
\right]
\label{C-closed-bis}
\end{eqnarray}
from which we can deduce the zero mode part of the boundary to be
\begin{eqnarray}
|B(E,F)\rangle_{z.m.}
&=&
%-\frac{T '_{25}}{2}
%N
%2^{-\frac{D}{4}} ~\sqrt{\det L} ~(\det  {\cal G})^{\frac{1}{4}}
|k_\mu=0\rangle 
\nonumber\\
&&\otimes
\sum_s e^{ i \pi \sum_{i < j} m^i \hat F_{i j} m^j} 
%\left[
\frac{  
e^{i \pi m^T 2 \theta} + e^{-i \pi m^T 2 \theta}}{2}% \right]
|n_i = \hat F_{i j} m^j, m^i=L_i s^i\rangle 
\nonumber\\
\label{B-state-bis}
\end{eqnarray}
The presence of Wilson lines does not change the tadpole condition
(\ref{tadpole}). 

\section{The Moebius amplitude.}
\label{moebius-sect}
Let us now check the result on the form of the crosscap and annulus
from the explicit computation of the Moebius amplitude while fixing
some details.

The Moebius amplitude is given by
\begin{equation}
Z_{M}=2*\int \frac{d \tau}{ 2 \tau} ~ Tr_o\left( \frac{\Omega}{2} e^{-2\pi \tau H_o}\right)
\end{equation}
where the trace is taken also over the Chan Paton factors and
the factor $2$ takes into account the two possible
orientations.
Given the previous case of two stacks we have a contribution only from
the strings connecting the two stacks since only these strings are
mapped into themselves 
by the worldsheet parity and we get
\begin{eqnarray}
Z_{M}&=& 
%N_1~ \int \frac{d \tau}{ 2 \tau}
%\int \frac{d^{D-d} k^\mu}{ (2 \pi)^{D-d}} 
%(2\pi)^{D-d} \delta^{D-d}(0) 
%e^{-2\pi \tau  \alpha' (k^\mu)^2} 
%\nonumber\\
%&&\times
%\sum_{n\in Z^{d}} \left[
%e^{ -2 \pi \tau (n+2 L \theta)^T L^{-1} {\cal G}^{-1}  L^{-1} (n+2 L \theta) }
%+ 
%e^{ -2\pi \tau (n-2 L \theta)^T L^{-1} {\cal G}^{-1}  L^{-1} (n-2 L \theta) }
%\right]
%(-)^{n_1 n_2+ \dots+ n_{2 r-1} n_{2 r}}
%\nonumber\\
%&&\times
%\frac{ e^{ 2 \pi \tau} }
% { \left( \prod \left(1- (-)^n e^{- 2\pi\tau n}\right)  \right)^{D-2}}
%\nonumber\\
%&=&
N_1
\int \frac{d \tau}{2\tau}
\frac{V_{nc}}{ (2 \alpha' \tau)^{\frac{D-d}{2}}}
\frac{ e^{ - \pi \tau \frac{26-D}{12} } e^{i \pi \frac{D-2}{24}} }
{  \left( f_1(i e^{-\pi\tau })\right)^{D-2} }
\nonumber\\
&&\times
\sum_{n\in Z^{d}} \left[
e^{ -2 \pi \tau (n+2 L \theta)^T L^{-1} {\cal G}^{-1}  L^{-1} (n+2 L \theta) }
+ 
e^{ -2\pi \tau (n-2 L \theta)^T L^{-1} {\cal G}^{-1}  L^{-1} (n-2 L \theta) }
\right] (-)^{n_1 n_2+ \dots+ n_{2 r-1} n_{2 r}}
\nonumber\\
\label{M-step1}
\end{eqnarray}
where $N_1 (-)^{n_1 n_2+ \dots+ n_{2 r-1} n_{2 r}}$ 
is the contribution from the Chan-Paton factors.
In particular the momentum dependent contribution is due to the fact
that
$\Lambda_2^T(n_1,n_2)=(-)^{n_1n_2}\Lambda_2(n_1,n_2)$.
Explicitly for a generic operator $O(p)$ commuting with $p$
we have
\begin{eqnarray}
&&
\langle J I' |
~{}_p\langle\frac{n_1}{\sqa L},\frac{n_2}{\sqa L} |
~\left(\Lambda^\dagger_{L}\right)_{ I' J}(n_1,n_2) 
~O(p)
\Omega
~\Lambda_{L; I' J}(n_1,n_2) 
~|\frac{n_1}{\sqa L},\frac{n_2}{\sqa L} \rangle_p
~|J I'\rangle
\nonumber\\
&&
=
(2\pi\sqa)^2  O(n)
~\left(\Lambda^\dagger_{L}\right)_{ I' J}(n_1,n_2) 
~\Lambda_{L; I' J}(n_1,n_2) 
\langle J I' |
~|I J'\rangle
\nonumber\\
&&
=
(2\pi\sqa)^2 O(n)
~tr\left(
\Lambda_{L}^\dagger(n_1,n_2) 
~\Lambda_{L}^T(n_1,n_2)
\right)
=
(2\pi\sqa)^2 O(n) (-)^{n_1n_2}
\end{eqnarray}
where the index $I$ belongs to the first stack and $J'$ to the mirror
one as the convention we introduced in section \ref{Omega-sect}.

The next step is to perform the Poisson resummation taking care of the
extra signs due to the momentum dependent Chan Paton factors.
To simplify the explanation of the computation we consider the
case of a factorized torus even if the comutation works the same way
in the non factorized case. 
For example for the first torus in the factorizes case we find
\begin{eqnarray}
&&\sum_{(n_1,n_2)\in Z^{2}} 
e^{ -2 \pi \tau (\frac{n}{2}+2 \theta)^T  {\cal G}_{(1)}^{-1} (\frac{n}{2}+2 \theta) }
(-)^{n_1 n_2}
=
\nonumber\\
&&\hspace{6em}=
\frac{1}{\sqrt{\det 2 \tau {\cal G}_{(1)}^{-1}}}
\sum_{(u^1,u^2)\in Z^{2}} 
e^{ - \pi \frac{1}{2\tau} u^T  {\cal G}_{(1)}^{-1} u }
e^{i 4\pi u^T \theta}
\left[
1+ e^{i \pi u^1}+ e^{i \pi u^2}- e^{i \pi (u^1+u^2)}
\right] 
\nonumber\\
&&\hspace{6em}=
\frac{ \sqrt{ \det {\cal G}_{(1)}} }{2 \tau}
\sum_{(u^1,u^2)\in Z^{2}} 
e^{ - \pi \frac{1}{2\tau} u^T  {\cal G}_{(1)}^{-1} u }
e^{i 4\pi u^T \theta}
\times 2 e^{i \pi u^1 u^2}
\end{eqnarray}
Notice that without the phase $(-)^{n_1 n_2}$ the terms in the square
brackets would give an overall factor of
$4\delta_{u_1,even}\delta_{u_2,even}$ and this would destroy the
tadpole cancellation since the Moebius would ahve the wrong normalization.

Given the previous result we can now proceed with the Poisson
resummation and the modular transformation to obtain
\begin{eqnarray}
Z_{M}&=& 
N_1
\frac{V_{nc}}{ \sqrt{\alpha'}^{D-d}}
\int \frac{d \tau}{2 \tau}
\frac{1}{(2\tau)^{(D-d)/2}}  
\left(\det 2 \tau  {\cal G}^{-1} \right)^{-1/2}
\nonumber\\
&&
\sum_{u\in Z^{d}} e^{ - \pi ~u^T \frac{{\cal G}}{2 \tau} ~u }
\frac{ e^{-4 i \pi u^T  \theta} + e^{+4 i \pi u^T  \theta} }{2}
~2^r
~(-)^{u^1 u^2+\dots u^{2r-1}u^{2r}}
\times
\frac{  \left(\tau \right)^{ \frac{D-2}{2} } 
        e^{ - \pi \tau \frac{26-D}{12} }  }
{  \left( f_1(e^{-\pi\frac{1}{\tau} })\right)^{D-2} }
\nonumber\\
&=&
{ N_1~\sqrt{\det L}}
\frac{V_{nc} ~\sqrt{ \det {\cal G} } }{2 \alpha'^{\frac{D-d}{2}}}
e^{i \pi \frac{D-2}{24}}
\int \frac{d \tau}{  \tau^2}
\sum_{u\in Z^{d}} e^{ -\pi u^T \frac{{\cal G}}{2\tau} u }
(-)^{\sum_{i<j} 2 B_{i j} u^i u^j}
\frac{ e^{-4 i \pi u^T \theta} + e^{+4 i \pi u^T \theta} }{2}
\nonumber\\
&&\times
\frac{  e^{ - \pi \tau \frac{26-D}{12} }  }
{  \left( f_1(i e^{-\frac{\pi}{4\tau} })\right)^{D-2} }
\nonumber\\
\label{M-step2}
\end{eqnarray}
where we have used $\det L = 2^{2 r}$.
Changing variable to $t=\frac{1}{4\tau}$ we get finally
\begin{eqnarray}
Z_M
&=&
e^{i \pi \frac{D-2}{24}}
~{ N_1~\sqrt{\det L}}
~\frac{V_{nc}  }{2 \alpha'^{\frac{D-d}{2}}}
~\sqrt{ \det {\cal G} }
\int d t
\sum_{u\in Z^{d}} e^{ -\pi u^T \frac{{\cal G}}{2\tau} u }
(-)^{\sum_{i<j} 2 B_{i j} u^i u^j}
\frac{ e^{-4 i \pi u^T  \theta} + e^{+4 i \pi u^T  \theta} }{2}
\nonumber\\
&&\times
\frac{  e^{ - \pi \tau \frac{26-D}{12} }  }
{  \left( f_1(i e^{-\frac{\pi}{4\tau} })\right)^{D-2} }
\nonumber\\
\label{M-step3}
\end{eqnarray}
which matches precisely the closed string computation
\begin{equation}
Z_M =\langle C(E) | 
\frac{\alpha' \pi }{2} \int_0^\infty d l~ 
e^{-\pi l (L_0+\tilde L_0  -2)}
\delta_{L_0,\tilde L_0}
| B(E,F)\rangle
~.
\end{equation}

\section{Conclusions.}
We have shown that in order to cancel the tadpole on an orbifold
compactification with a non trivial discrete half integer $B_{i j}$ 
turned on 
it is necessary to turn an equal constant magnetic field on the branes
$\hat F_{i j}=B_{i j}$: 
this is the only consistent value for the background field strength in
those directions allowed by tadpole cancellation.
Since $\hat F_{i j}$ is not integer is cannot be a field strength of a
$U(1)$ bundle but  
it must be defined in a non trivial bundle with a non abelian
structure group on a torus.
The string then describes the fluctuations around this background,
these fluctuations have not the usual spectrum because of
mechanism like a Scherk-Schwarz reduction 
and this explains in a very intuitive way why a
configuration with  $B$ rank $rk(B)=r$ reduces the rank of the gauge
group as $SO(2^{D/2})\supset SO(2)^r \otimes SO( 2^{D/2-r})\rightarrow
SO( 2^{D/2-r})$.

%\vskip0.5cm
%{\bf Acknowledgments}

%\noindent
%The author thanks ... for discussions.

\end{document}